\documentclass[%
 reprint,
superscriptaddress,
 amsmath,amssymb,
aps,
pra
]{revtex4-2}
\usepackage{gensymb}
\usepackage{graphicx}   
\draft 
\usepackage{xcolor}
\usepackage{braket}

\begin{document}

\title{Highly crystalline superconducting TiN resonators grown on thermally reconstructed sapphire} 

\author{Thomas J. Smart}
\affiliation{Peter Grünberg Institute (PGI-9), Forschungszentrum J\"ulich and JARA J\"ulich-Aachen Research Alliance, 52425 J\"ulich, Germany}

\author{Marc Neis}
\affiliation{Institute for Functional Quantum Systems (PGI-13), Forschungszentrum J\"ulich, 52425 J\"ulich, Germany}
\affiliation{RWTH Aachen University, 52074 Aachen, Germany}

\author{Janine Lorenz}
\affiliation{Peter Grünberg Institute (PGI-3), Forschungszentrum J\"ulich and JARA J\"ulich-Aachen Research Alliance, 52425 J\"ulich, Germany}
\affiliation{Institute of Experimental Physics IV A, RWTH Aachen University, 52074 Aachen, Germany}

\author{Marcello P. Guardascione}
\affiliation{Institute for Functional Quantum Systems (PGI-13), Forschungszentrum J\"ulich, 52425 J\"ulich, Germany}
\affiliation{RWTH Aachen University, 52074 Aachen, Germany}

\author{Roudy Hanna}
\affiliation{Peter Grünberg Institute (PGI-9), Forschungszentrum J\"ulich and JARA J\"ulich-Aachen Research Alliance, 52425 J\"ulich, Germany}
\affiliation{Institute for Functional Quantum Systems (PGI-13), Forschungszentrum J\"ulich, 52425 J\"ulich, Germany}
\affiliation{RWTH Aachen University, 52074 Aachen, Germany}

\author{Michael Schleenvoigt}
\affiliation{Peter Grünberg Institute (PGI-9), Forschungszentrum J\"ulich and JARA J\"ulich-Aachen Research Alliance, 52425 J\"ulich, Germany}

\author{Yuan Gao}
\affiliation{Institute for Functional Quantum Systems (PGI-13), Forschungszentrum J\"ulich, 52425 J\"ulich, Germany}
\affiliation{RWTH Aachen University, 52074 Aachen, Germany}

\author{Joscha Domnick}
\affiliation{Peter Grünberg Institute (PGI-9), Forschungszentrum J\"ulich and JARA J\"ulich-Aachen Research Alliance, 52425 J\"ulich, Germany}

\author{Benjamin Bennemann}
\affiliation{Peter Grünberg Institute (PGI-10), Forschungszentrum J\"ulich and JARA J\"ulich-Aachen Research Alliance, 52425 J\"ulich, Germany}

\author{Abdur Rehman Jalil}
\affiliation{Institute for Experimental Physics III, University of W\"urzburg, 97074 W\"urzburg, Germany}
\affiliation{Peter Grünberg Institute (PGI-10), Forschungszentrum J\"ulich and JARA J\"ulich-Aachen Research Alliance, 52425 J\"ulich, Germany}

\author{Jin Hee Bae}
\affiliation{Peter Grünberg Institute (PGI-9), Forschungszentrum J\"ulich and JARA J\"ulich-Aachen Research Alliance, 52425 J\"ulich, Germany}

\author{Harsh Bhardwaj}
\affiliation{Institute for Functional Quantum Systems (PGI-13), Forschungszentrum J\"ulich, 52425 J\"ulich, Germany}
\affiliation{RWTH Aachen University, 52074 Aachen, Germany}

\author{F. Stefan Tautz}
\affiliation{Institute of Experimental Physics IV A, RWTH Aachen University, 52074 Aachen, Germany}
\affiliation{Peter Grünberg Institute (PGI-3), Forschungszentrum J\"ulich and JARA J\"ulich-Aachen Research Alliance, 52425 J\"ulich, Germany}

\author{Felix Lüpke}
\affiliation{Peter Grünberg Institute (PGI-3), Forschungszentrum J\"ulich and JARA J\"ulich-Aachen Research Alliance, 52425 J\"ulich, Germany}
\affiliation{Institute of Physics II, University of Cologne, 50937 Cologne, Germany}

\author{Detlev Grützmacher}
\affiliation{Peter Grünberg Institute (PGI-9), Forschungszentrum J\"ulich and JARA J\"ulich-Aachen Research Alliance, 52425 J\"ulich, Germany}

\author{Rami Barends}
\affiliation{Institute for Functional Quantum Systems (PGI-13), Forschungszentrum J\"ulich, 52425 J\"ulich, Germany}
\affiliation{RWTH Aachen University, 52074 Aachen, Germany}

\author{Pavel A. Bushev}
\affiliation{Institute for Functional Quantum Systems (PGI-13), Forschungszentrum J\"ulich, 52425 J\"ulich, Germany}

\author{Peter Schüffelgen}
\email{p.schueffelgen@fz-juelich.de}
\affiliation{Peter Grünberg Institute (PGI-9), Forschungszentrum J\"ulich and JARA J\"ulich-Aachen Research Alliance, 52425 J\"ulich, Germany}


\def\thefootnote{*}\footnotetext{These authors contributed equally to this work}\def\thefootnote{\arabic{footnote}}

\date{\today}

\begin{abstract}
High-quality crystalline growth of a thin film on sapphire requires sufficient substrate preparation, often achieved via aggressive chemical cleaning. Direct thermal reconstruction of the sapphire substrate via a CO$_2$ laser beam may provide an alternative method to prepare the substrate for epitaxy without the use of any chemical processing. In this work, we demonstrate that the thermal annealing of sapphire into its ($\sqrt{31}$$\times$$\sqrt{31}$)\,R$\pm$9\degree reconstruction is a valid alternative preparation technique for sapphire substrates. TiN films grown via plasma-assisted molecular beam epitaxy upon these substrates exhibit greater crystallinity than those grown on chemically-cleaned sapphire substrates. Superconducting resonators fabricated from these films exhibit similar performance, with many possessing internal quality factors at single photon levels greater than 10$^6$ for both substrate preparation methods.
\end{abstract}

\pacs{}

\maketitle 

\section{Introduction}
While Nb-based technologies have a long history in superconducting electronics \cite{Kuroda1988}, Al/AlO$_x$ has become a widely adopted material system for superconducting quantum circuits \cite{Barends2014,Nakamura1999}. Alongside these established platforms, alternative superconducting materials have been investigated for decades with the aim of improving device performance and expanding fabrication capabilities \cite{Wang1994,Face1987,Place2021}. Superconducting transition metal nitrides (TMNs) have attracted significant attention because of their desirable properties, such as a large superconducting gap, high critical temperature, $T_c$, and resistance to oxidation and chemical etching \cite{Sundgren1985,Lengauer2015}.
\par
Titanium Nitride (TiN) is a metallic compound that is readily used within the semiconductor industry due to its desirable properties, such as: high electrical conductivity, compatibility with existing CMOS technology and chemical stability \cite{Mauersberger2021, Briggs2016}. Despite its lower $T_c$ and superconducting gap size, TiN offers a comparatively simple phase-stability landscape for high-temperature epitaxial growth, compared to other popular superconducting TMNs, like the Nb-N system \cite{Wright2021,Takiguchi2025}. In recent years, TiN has been shown to be a promising material for passive superconducting components, with the internal quality factors ($Q_i$) of prior superconducting co-planar waveguide (CPW) resonators, exceeding $10^6$ at single photon levels for films grown on silicon and sapphire substrates \cite{Richardson2020,Vissers2010,Ohya2014,Shearrow2018}. Commonly reported methods for the chemical cleaning of sapphire substrates include organic solvent cleaning, typically using acetone and isopropanol, as well as more aggressive treatments using Piranha solution or hydrofluoric acid (HF) to remove surface contaminants and obtain a clean substrate surface \cite{Zhang2013}.

An alternative substrate preparation method is presented through the thermal reconstruction of the surface of the substrate \cite{Braun2020}. In the case of sapphire substrates, \textit{in-situ} direct heating by a CO$_2$ laser substrate heater, such as those available with thermal laser epitaxy (TLE) chambers, provides a rapid method to thermally prepare the substrate for further epitaxy. Crucially, this rapid thermal preparation of the substrate is performed without any aggressive $ex-situ$ chemical preparation \cite{SmartSiTLE2024,Smink2024}. $C$-plane sapphire substrates that have been reconstructed into the characteristic ($\sqrt{31}$$\times$$\sqrt{31}$)\,R$\pm$9\degree (R31) reconstruction have been shown to enable the high-quality crystalline growth of various materials with pristine interface quality and exemplary electronic properties \cite{MajerTa2024,MajerRu2024,KimCarbon2023,KimNitridesTLE2025, SmartSiTLE2024}. Given this platform, an intriguing opportunity is presented to observe if thermal reconstruction of the sapphire substrate impacts the performance of a simple superconducting device, such as a CPW resonator, that has been epitaxially grown upon the substrate. This can then be compared to a sister device grown and fabricated on sapphire substrates that have been prepared with aggressive chemicals.
\par
Within this work, we epitaxially grow highly crystalline TiN (111) films on both chemically-cleaned $c$-plane sapphire and thermally reconstructed R31 sapphire via plasma-assisted Molecular Beam Epitaxy (MBE). It was observed that TiN films grown on reconstructed sapphire exhibit a greater degree of crystallinity compared to those grown on sapphire prepared via aggressive acids. By etching the films using resist-based lithography, we manufacture superconducting CPW resonators from these films. For both preparation methods, we observe $Q_i$ values exceeding 10$^6$ at single photon values, indicating that thermal reconstruction provides a rapid and simple alternative to substrate preparation via aggressive acids, without significantly affecting superconducting properties of the films nor the performance of a corresponding superconducting device.  

\section{Methods}
\subsection{Growth}
Pre-diced 10$\times$10\,mm$^2$ pieces of $c$-plane sapphire ($\alpha$-Al$_2$O$_3$) supplied by CrysTec GmbH were used as the substrate during this work. Before any further processing of the substrates in a cleanroom environment, all substrates were first cleaned with an acetone and isopropanol (IPA) bath of 5\,min each with sonication, before further chemical or thermal treatment to avoid equipment contamination. A selection of these substrates were then chemically processed further in a Piranha solution mix of 2:1 H$_2$SO$_4$:H$_2$O$_2$ for 10\,min, followed by a dip in a 1\% solution of HF for 1\,min. A final cleaning step of a 10\,min dip in deionised (DI) water followed by a 2\,min dip in isopropanol was performed to remove any volatile products and ensure a clean surface. These chemically-processed Al$_2$O$_3$ substrates were then loaded into an ultra-high vacuum (UHV) preparation chamber, where they were baked at 900\,\degree C for 45\,min to prepare an epi-ready surface without forming a surface reconstruction \cite{BRUCKER2025}. After baking, these substrates were loaded into the MBE chamber for deposition. Sapphire substrates cleaned in this manner will be hereon referred to as `Bare' sapphire or B-Al$_2$O$_3$.
\par
Another portion of the acetone/IPA cleaned substrates were prepared thermally by loading them into a TLE system, which operates at ultra-high vacuum (UHV) conditions with a chamber pressure $p \sim 1 \times 10^{-9}$\,mbar during laser irradiation. \cite{Braun2019,SmartThesis,Smart2021}. This system is equipped with a CO$_2$ laser substrate heater with a laser beam diameter of 5\,cm, capable of directly heating the Al$_2$O$_3$ substrates to extremely high temperatures \cite{Hanna2025, KimCarbon2023,Smink2024}. This method provides a far greater range of substrate temperatures without the risk of contamination caused by indirect heating via wire heaters or tube furnaces \cite{Braun2020}. Under these conditions, the surface of the Al$_2$O$_3$ substrates reforms into the R31 reconstruction described previously. The Al$_2$O$_3$ substrates were held at 1700\,\degree C for 200\,s, forming a terraced structure with a step height of $\sim$\,0.4\,nm, corresponding to two atomic layers of the Al$_2$O$_3$ unit cell as previously described in prior literature \cite{Smink2024}. The presence of a R31 surface reconstruction was verified via Reflection High-Energy Electron Diffraction (RHEED) imaging performed after heating, in which a clear double-step R31 reconstruction is observed. These images are shown within Appendix \ref{RHEED_Appendix}.

\begin{figure}[!h]
    \includegraphics[width =\linewidth]{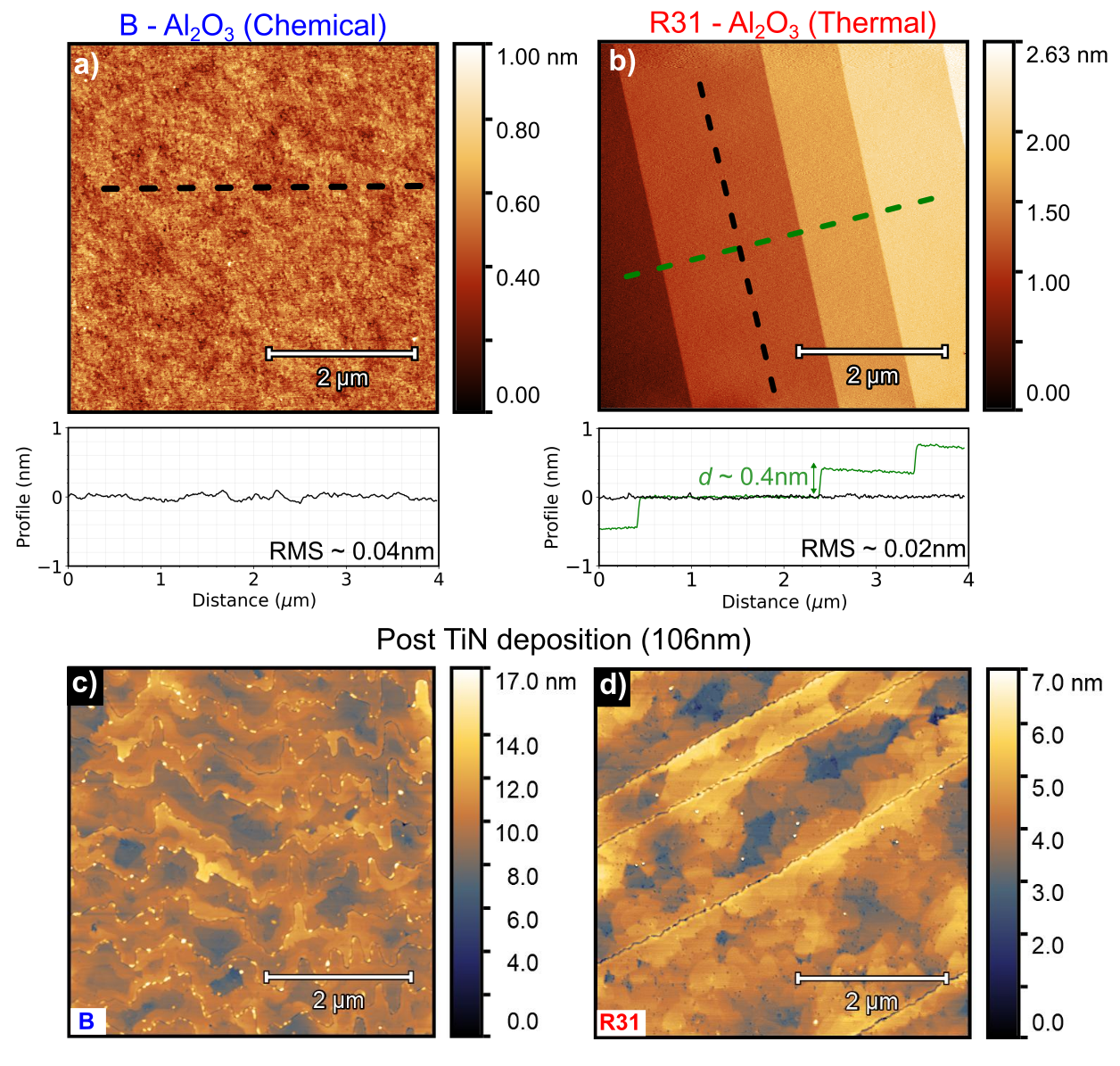}
    \caption{a),b) AFM topographs of $c$-plane sapphire substrates, after chemical cleaning with Piranha solution and 1\% HF  (B-Al$_2$O$_3$) and after thermal annealing to form the ($\sqrt{31}$ $\times$ $\sqrt{31}$)\,R$\pm$9\degree reconstruction at T$_{\mathrm{sub}}$ = 1700\,\degree C for 200\,s (R31-Al$_2$O$_3$). Height profiles along indicated paths are shown underneath each AFM topograph with the calculated RMS roughness noted for each substrate. c),d) AFM topographs of $d$ = 106\,nm TiN films grown at 1150\,\degree C.}
    \label{fig1b} 
\end{figure}

The change in the surface morphology during surface reconstruction compared to chemically cleaned Al$_2$O$_3$ is shown in Fig.~\ref{fig1b}a and Fig.~\ref{fig1b}b, respectively. Parallel to each terrace on the reconstructed sapphire, the root-mean-square (RMS) roughness is noticeably lower compared to the bare sapphire substrate. The flat and well-ordered surface morphology provides a favorable platform for epitaxy with atomically sharp interfaces compared to unreconstructed Al$_2$O$_3$, which recent literature has shown does not provide a uniform and smooth (1 $\times$ 1) surface \cite{Hutner-Reisch2026}. The use of reconstructed R31-Al$_2$O$_3$ within physical vapor deposition and epitaxy promotes the growth of highly crystalline thin films with exemplary crystal properties, as previously demonstrated for various material systems \cite{KimNitridesTLE2025,SmartSiTLE2024,MajerRu2024,MajerTa2024}. As confirmed by the RHEED images in Appendix \ref{RHEED_Appendix} and in prior literature, once the R31 reconstruction has formed, it remains stable in air, even after months in atmospheric conditions \cite{Smink2024}. Consequentially, $ex-situ$ transfer of the reconstructed sapphire between the TLE chamber and any separate growth chamber did not effect the quality of substrate, provided that any volatile products like water vapor were removed prior to epitaxy \cite{Smink2024}. Sapphire substrates prepared in this manner will be henceforth referred to as R31-Al$_2$O$_3$.

\par

A selection of B-Al$_2$O$_3$ and R31-Al$_2$O$_3$ substrates were then simultaneously loaded into a MBE chamber, where they were first baked at 200\,\degree C for 45\,min to remove any lingering water vapor or other adsorbents. Prior to growth, the chamber operated under UHV conditions with $p \sim 1 \times 10^{-10}$\,mbar.
During growth within the MBE chamber, an atomic nitrogen (N) atmosphere was provided by a Radio Frequency (RF) plasma source with an RF power of 400\,W and a N flow rate of 1.25\,sccm, providing a chamber pressure of 1.5 $\times$ $10^{-4}$\,mbar during growth. Titanium (Ti) was provided via an e-beam evaporator held at a partial pressure of 1 $\times$ $10^{-8}$\,mbar, resulting in a TiN growth rate of $\sim$ 0.4\,\AA/s. Films with a thickness $d$ = 106\,nm of TiN were grown at 1150\,\degree C. To ensure the stoichiometry of the deposited TiN is as close to the desired 1:1 atomic ratio of Ti:N as possible, the deposited film was held at 1100\,\degree C for an hour within the atomic N atmosphere after growth was complete. This allowed for any volatile materials incorporated during growth to leave the TiN film whilst fully nitriding the remaining sample. The sample was then cooled back down to room temperature over the period of half an hour within the atomic N atmosphere. The presence of this atmosphere during cool down was designed to reduce the diffusion of N out of the TiN at high temperatures, which may negatively impact the superconducting properties of the grown TiN film \cite{Takiguchi2025}. A second set of TiN films with $d$ = 17\,nm were also grown on both R31-Al$_2$O$_3$ and B-Al$_2$O$_3$, under the same growth parameters described previously for imaging via Transmission Electron Microscopy (TEM). All other measurements were performed on the $d$ = 106\,nm TiN films.

\subsection{Design and Fabrication of Resonators}
\begin{figure}[!h]
    \includegraphics[width =\linewidth]{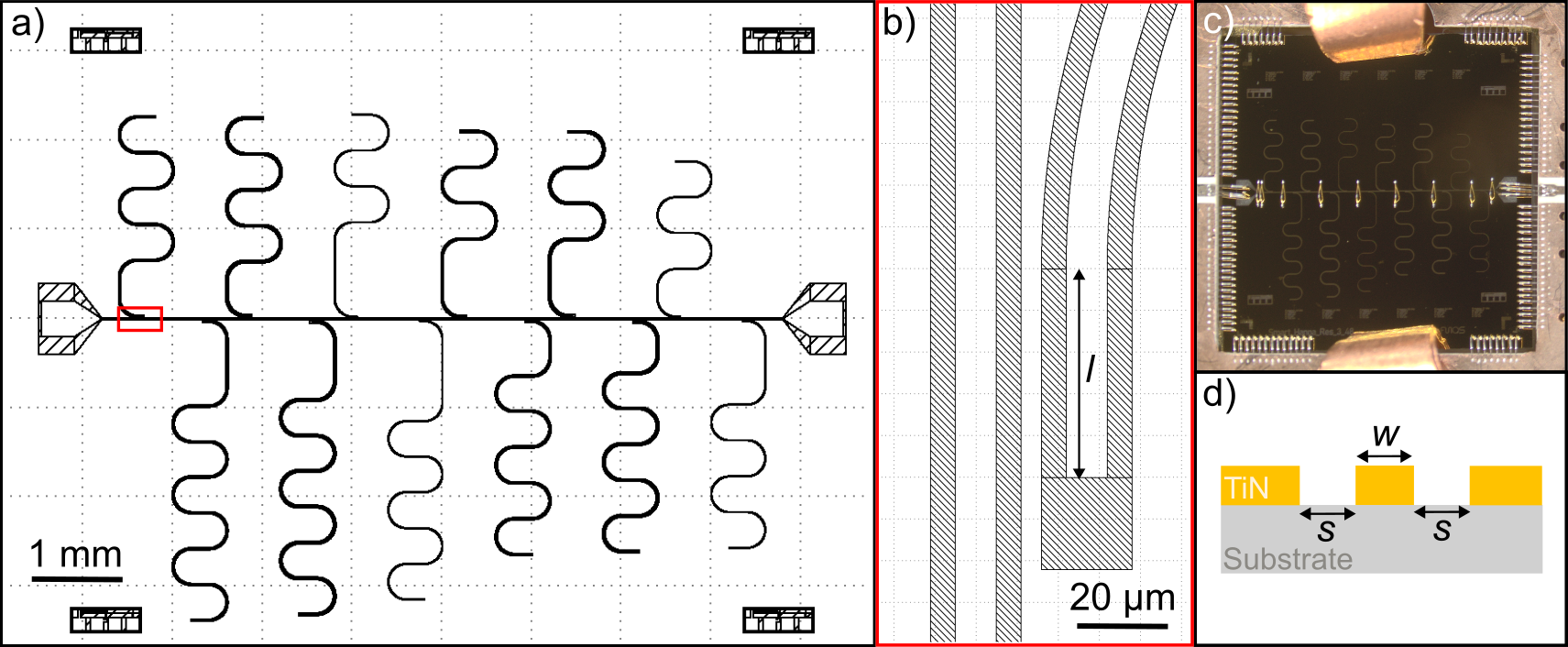}
    \caption{a) Drawing of the $\lambda$/4 CPW resonator design used in this work. The four rectangular structures, each placed at the corners of each chip are four-point probe structures to measure the superconducting transition as a function of temperature (see Appendix \ref{App_SC-Properties}). b) Zoom in of resonator design showing the coupling between the central feedline and a resonator, with the coupler length, $l = 50\,\mu m$ indicated. This value remained fixed for all resonators. c) Photograph of 10 $\times$ 10\,mm fabricated TiN resonator chip after bonding. d) Cross-sectional diagram of CPW resonator, where the width of the central conductor, $w$ and the gap size, $s$ are indicated.}
    \label{ResonatorDesign} 
\end{figure}

Once growth was completed, the $d$ = 106\,nm TiN films were fabricated into $\lambda$/4 CPW resonators using a standard optical resist lithography process and dry etching. The samples were spin-coated with ECI-3012 (positive) photo-resist, soft baked at 90\,\degree C for 1\,min and then patterned using direct laser writing via a Maskless Aligner (MLA) device with a dose of 130\,mJ/cm$^2$. The samples were then post-exposure baked at 110\,\degree C for 1\,min before developing the exposed resist via a 65\,s dip in AZ-326 metal-ion free (MIF) solution. The TiN films were then patterned via reactive ion etching (RIE) using SF$_6$ gas with a flow rate of 30\,sccm, an RF power of 25\,W and an ICP power of 100\,W. Once the samples had been etched, the remaining resist was removed with acetone and isopropanol dips of 10\,min each. 
A drawing of the $\lambda$/4 CPW resonator design used in this work is given in Fig. \ref{ResonatorDesign}a. 

The structured CPW resonators were designed to have their resonance frequencies lie within the range of 4-8\,GHz, consisting of a range of different structures sizes and coupling to a central readout line. For all resonators, the coupler length, $l$, as shown in Fig. \ref{ResonatorDesign}b, was fixed at 50\,$\mu$m. The central conductor width, $w$ and the gap size, $s$ between the central conductor and the remaining TiN film was varied for each resonator. A full list of resonator parameters is given in Appendix \ref{App_ResonatorParameters}.

\subsection{Cryogenic Measurement Setup}
The electronic properties of the material were probed in a BlueFors SD cryogenic dilution refrigerator, which cooled the samples down to $\sim$ 20\,mK. Using the build-in heaters the setup can be used at a range of temperatures to measure the resistance via four-point probe structures determining DC properties such as resistivity, $\rho$ and $T_c$. Additionally, the cryogenic experiment was used to measure the electronic properties of any device at frequencies ranging from 3.5\,GHz to 8.5\,GHz. These chips were connected to custom made PCBs via Aluminium wire-bonding, which have SMA connectors for readout. A Keysight VNA (N5222B), in series with a total of 60\,dB of attenuation on the various temperature stages, probed the S-parameters of the samples.

\section{Results}
\subsection{Growth and Film Quality}
\begin{figure}[!h]
    \includegraphics[width =\linewidth]{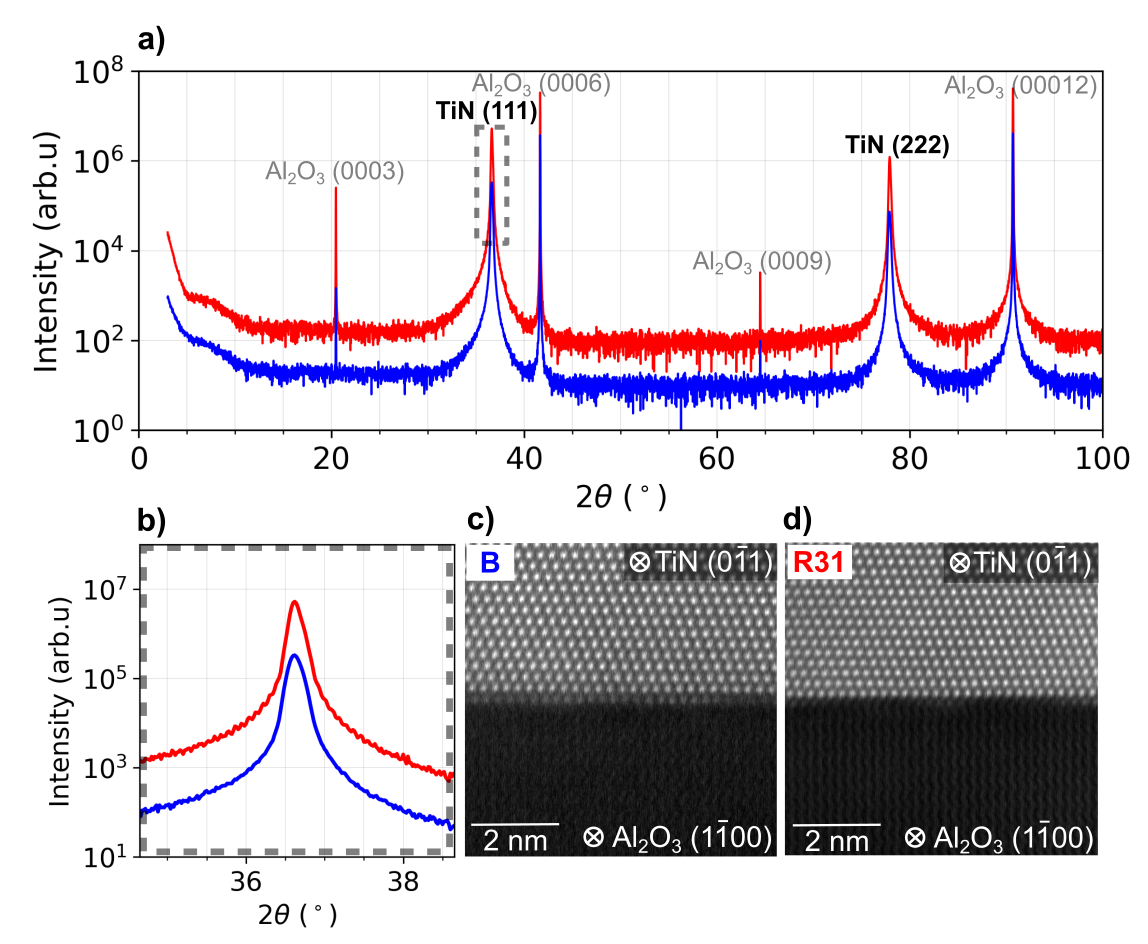}
    \caption{a) Comparison of XRD spectra for $d$ = 106\,nm TiN films grown on reconstructed (R31-Al$_2$O$_3$ - Red) and bare sapphire (B-Al$_2$O$_3$ - Blue), where the relevant features have been indicated. The R31-Al$_2$O$_3$ curve has been vertically displaced by a factor of 10 for clarity. b) Zoom-in on the TiN (111) peak from the XRD spectra. c),d) Transmission Electron Microscopy (TEM) images of the TiN/Al$_2$O$_3$ interface of $d$ = 17\,nm thick TiN films for B-Al$_2$O$_3$ and R31-Al$_2$O$_3$ respectively. The interface appears sharper for R31-Al$_2$O$_3$ compared to B-Al$_2$O$_3$.}
    \label{fig2a} 
\end{figure}

The TiN films grown on reconstructed (R31-Al$_2$O$_3$) and bare sapphire (B-Al$_2$O$_3$) were analysed with a variety of crystallography and imaging techniques. Firstly, the TiN films grown on R31-Al$_2$O$_3$ and B-Al$_2$O$_3$ have visually different morphologies. As shown in the Atomic Force Microscopy (AFM) micrographs in Fig.~\ref{fig1b}c and Fig.~\ref{fig1b}d, TiN grown on R31-Al$_2$O$_3$ demonstrates a smoother surface morphology with micron-scale grains that are aligned with the underlying terrace structure of R31-Al$_2$O$_3$. For both substrate preparation methods, the TiN films exhibit epitaxial growth with a face-centered cubic structure and (111) surface orientation. As shown in Fig.~\ref{fig2a}a, both films exhibit sharp and intense TiN (111) peaks. The fixed azimuthal relationship between the TiN (111) and Al$_2$O$_3$ poles is preserved for both films on R31-Al$_2$O$_3$ and B-Al$_2$O$_3$, indicating the same TiN/Al$_2$O$_3$ epitaxial registry in both films. Pole figures, shown with Appendix \ref{App_PoleFigures}, reveal a sixfold TiN pole distribution that is consistent with two 60$\degree$-related in-plane rotational variants of TiN (111).

The quality of the TiN/Al$_2$O$_3$ interface was examined under cross-sectional TEM on two sister TiN samples with $d$ = 17\,nm. Owing to the large lattice mismatch between TiN(111) and $c$-plane Al$_2$O$_3$ of approximately 8.5\%, epitaxial strain is relieved within the first few atomic layers due to the formation of interfacial misfit dislocations \cite{Rasic2017}. This results in the interface structure between TiN(111) and $c$-plane Al$_2$O$_3$ being fixed in the initial stages of growth \cite{Narayan2003}. Under identical deposition conditions, cross-sectional TEM performed on a $d$ = 17\,nm TiN film is therefore expected to be representative of the TiN/Al$_2$O$_3$ interface within thicker TiN films, although microstructural features within the film itself that evolve with film thickness, such as threading defects and grain morphology, may differ.
\par
No significant visual evidence of intermixing between the TiN and R31-Al$_2$O$_3$ substrate was observed for either sample. There is some indication of a smoother interface between the TiN and R31-Al$_2$O$_3$, but both substrates exhibit a low surface roughness along the imaged crystal axis. This reduced roughness of R31-Al$_2$O$_3$ is consistent with the AFM measurements performed in Fig.~\ref{fig1b} ans supports the reduced surface roughness of each terrace in the ($\sqrt{31}$ $\times$ $\sqrt{31}$)\,R$\pm$9\degree reconstruction relative to its chemically-cleaned counterpart \cite{Smink2024,BRUCKER2025}. 

\begin{figure}[!h]
    \includegraphics[width =\linewidth]{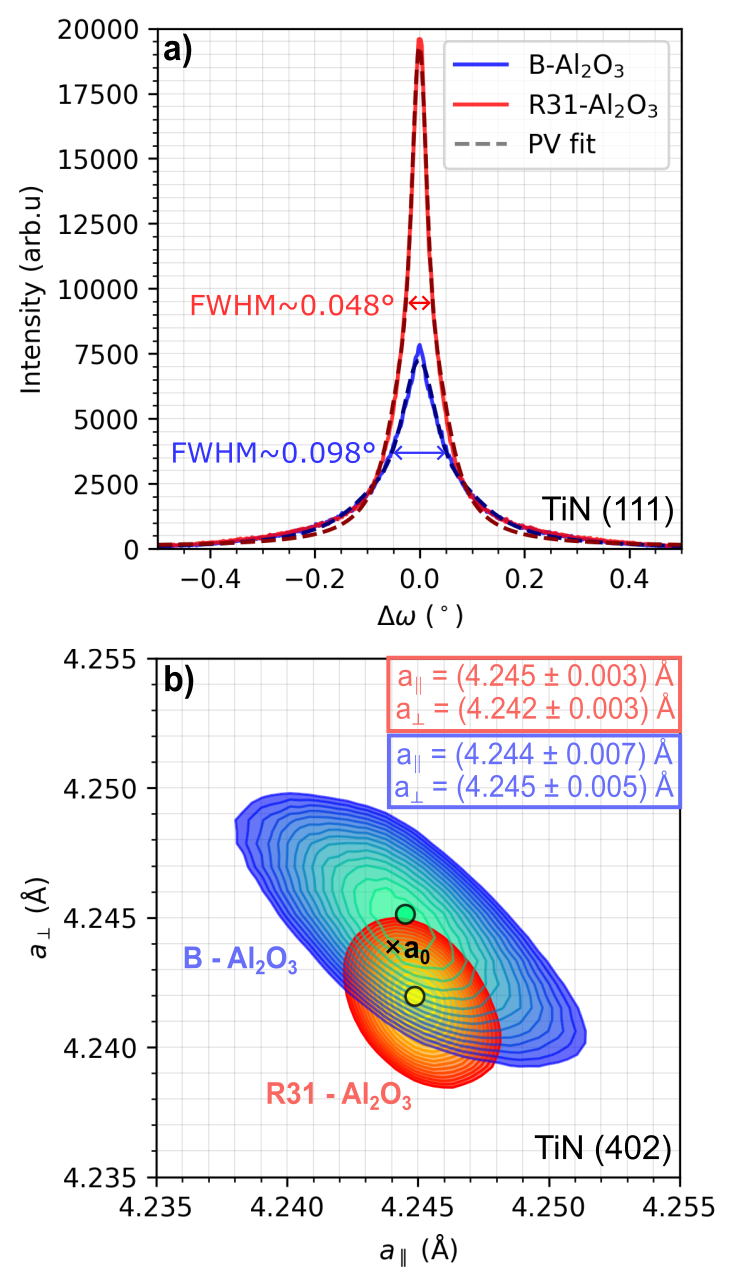}
    \caption{a) TiN (111) rocking curves for films with $d$ = 106\,nm grown on R31-Al$_2$O$_3$ and B-Al$_2$O$_3$. Both curves have been fit to a Pseudo-Voigt (PV) distribution (shown as dotted lines) to extract the FWHM values of 0.047\degree and 0.098\degree respectively. b) RSM plots of the off-axis TiN (402) peak showing the distribution of in-plane ($a_{\parallel}$) and out-of-plane ($a_{\perp}$) lattice constants for each film. Only intensity values greater than 50\% of the maximum value have been plotted for clarity. The literature value of the TiN lattice constants, ($a_0$) is indicated by a black cross whilst the peak intensity for each distribution is indicated by a colored dot. The average value for $a_{\parallel}$ and $a_{\perp}$ for each film is also indicated.}
    \label{RSM} 
\end{figure}

A smoother R31-Al$_2$O$_3$ surface would serve as a favorable platform for greater crystalline coherence, due to its greater crystal order and lower surface roughness \cite{Hutner-Reisch2026}. This is supported by rocking curve measurements around the TiN(111) peak for both $d$ = 106\,nm films, wherein the rocking curve for the TiN sample grown on R31-Al$_2$O$_3$ is noticeably more intense and sharper than the corresponding curve of TiN on B-Al$_2$O$_3$. When both curves are fit to a pseudo-Voigt dependence, shown as dotted lines in Fig.~\ref{RSM}a, the extracted full-width-half-maximum (FWHM) values for the TiN films on R31-Al$_2$O$_3$ and B-Al$_2$O$_3$ are 0.048\degree and 0.098\degree respectively.
\par

From XRD measurements, Reciprocal Space Maps (RSMs) were extracted to elucidate how the distribution of in-plane ($a_{\parallel}$) and out-of-plane ($a_{\perp}$) lattice constants of the TiN films are effected by the different substrate preparation methods. Examining the off-axis (402) reflections of the TiN films allows for information regarding $a_{\parallel}$ and $a_{\perp}$ to be directly extracted. In Fig.~\ref{RSM}b, the corresponding RSMs from the TiN(402) reflection are overlaid, where only intensity values greater than half of the maximum intensity are plotted. For both substrates, $a_{\parallel}$ and $a_{\perp}$ are very similar in value, as expected for a cubic crystal structure. For R31-Al$_2$O$_3$, The TiN film possesses $a_{\parallel}$ = (4.245 $\pm$ 0.003)\,\AA{} and $a_{\perp}$ = (4.242 $\pm$ 0.003)\,\AA. For TiN grown on B-Al$_2$O$_3$, the film exhibits $a_{\parallel}$ = (4.244 $\pm$ 0.007)\,\AA{} and $a_{\perp}$ = (4.245 $\pm$ 0.005)\,\AA.

The extracted lattice constants correspond to in-plane strains of $\epsilon_{\parallel}$ = (0.021 $\pm$ 0.064)\% and  $\epsilon_{\parallel}$ = (0.002 $\pm$ 0.165)\% for TiN on R31- and B-Al$_2$O$_3$ respectively, relative to bulk TiN (a = 4.244\,\AA{}) \cite{Catellani2017,Zhang2021}. The corresponding out-of-plane strains are $\epsilon_{\perp}$ = (-0.048 $\pm$ 0.072)\% and $\epsilon_{\perp}$ = (0.024 $\pm$ 0.118)\%.

Both in-plane and out-of-plane strain values are relatively small, suggesting that the film is largely relaxed. The contour shapes shown in Fig. \ref{RSM}b further support these interpretations; the elongated distribution for  B-Al$_2$O$_3$ lies along a single primary axis in the $a_{\parallel}$ - $a_{\perp}$ plane, reflecting correlated fluctuations between in-plane and out-of-plane lattice constants. This behavior is characteristic of microstrain broadening and mosaic disorder, where minor local lattice tilts and distortions contribute to the spread of values \cite{Moram2009,Calamba2019}. In contrast, the nearly circular contour for R31-Al$_2$O$_3$ indicates an isotropic distribution of lattice constants, consistent with a more uniform strain state and reduced defect density. The overall weaker intensity and larger spread for B-Al$_2$O$_3$ suggest a smaller coherent scattering volume, likely arising from higher dislocation density or reduced grain size, whereas the sharper, more intense distribution for R31-Al$_2$O$_3$ reflects improved crystalline coherence. When taken together, these results point to R31-Al$_2$O$_3$ providing a more favorable template for crystalline TiN growth, enabling reduced mosaicity and with more homogeneous strain relaxation compared to B-Al$_2$O$_3$.

\begin{figure}[!h]
    \includegraphics[width =\linewidth]{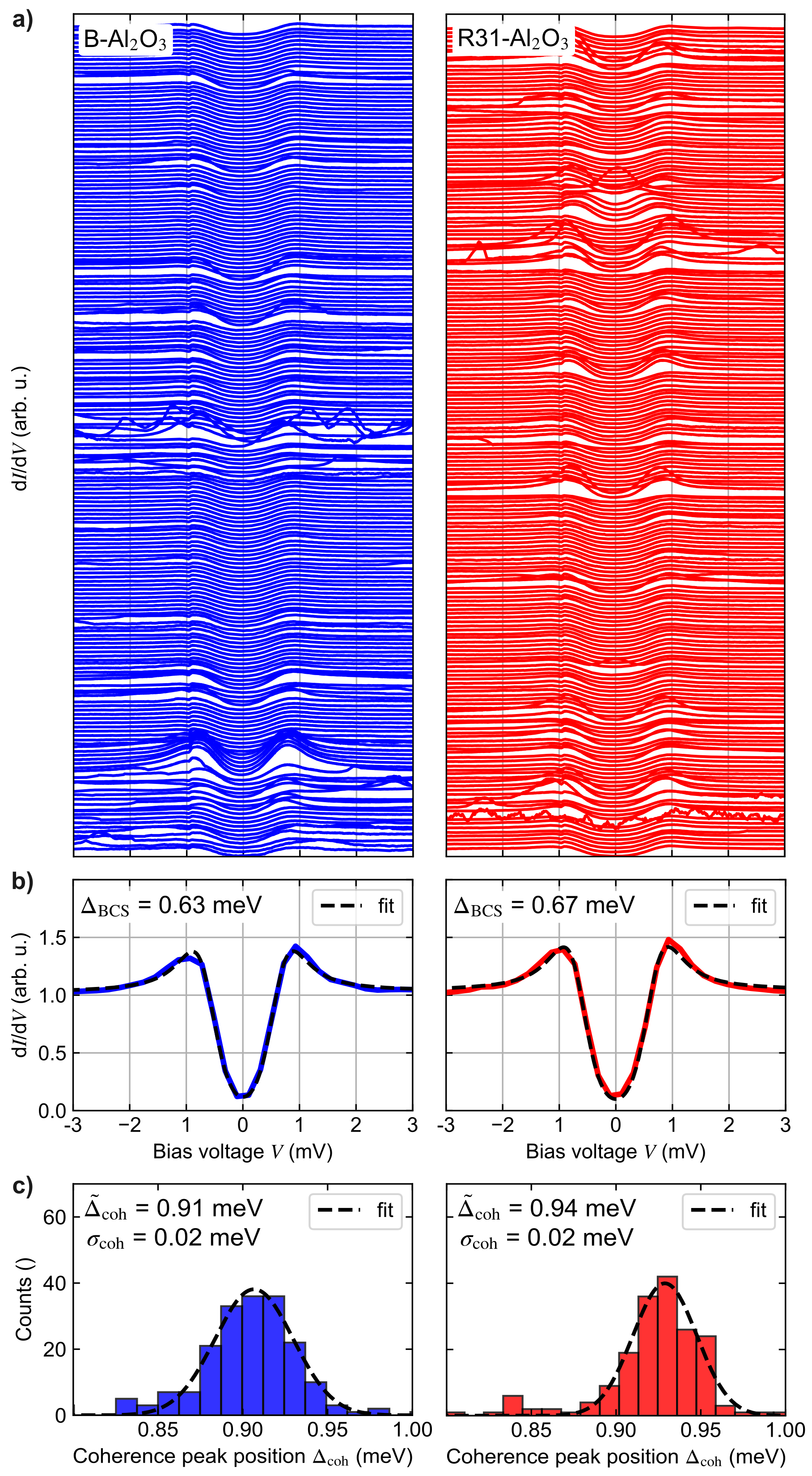}
    \caption{a) Waterfall plot of 200 differential tunneling conductance spectra recorded along a 424\,nm line across a representative area of the surface of each TiN film with $d$ = 106\,nm grown on B-Al$_2$O$_3$ and R31-Al$_2$O$_3$ (successive spectra are taken every 2.1\,nm, and are offset vertically). 
    b) Average of the 200 spectra shown in a) including BCS fits which yield order parameters of $\Delta_{\rm BCS}=0.63$ and 0.67\,meV as well as effective temperatures of $T_{\rm tip}=1.90$ and 1.91\,K for the samples B-Al$_2$O$_3$ and R31-Al$_2$O$_3$ respectively. c) Distribution of the energy at which the coherence peak maxima occur $\Delta_{\rm coh}$ in the spectra shown in panel a). The mean values and standard deviations of the distributions are determined from the shown Gaussian fits and indicated as insets. 
    Tunneling spectra were recorded at an STM temperature of 1.2\,K with a set point current of 1\,nA and a lock-in amplitude of 0.08\,mVpp at a modulation frequency of 819\,Hz.
    }
    \label{spectra} 
\end{figure}
\par
Both TiN films were measured to turn superconducting with $T_c$ = 5.08\,K and 5.11\,K for B-Al$_2$O$_3$ and R31-Al$_2$O$_3$ respectively. Via the use of fabricated four-point probe structures, $\rho$ of each film was measured at room temperature and at temperatures close to $T_c$. From these measurements, the Residual Resistance Ratio (RRR) value for each film was calculated. These were found to be broadly similar at 2.64 and 2.62 for B-Al$_2$O$_3$ and R31-Al$_2$O$_3$ respectively, supporting the comparable level of film quality between both samples observed in Fig.~\ref{fig2a}. 

Whilst it is evident that TiN grown at high temperatures on R31-Al$_2$O$_3$ exhibits greater crystallinity, it may not definitively translate to the superconducting properties, particularly since both can be considered to be disordered superconductors \cite{Kamlapure2013,Dynes1986}. To probe if the observed differences in film crystallinity translate to changes in the superconducting properties, we perform scanning tunneling microscopy/spectroscopy (STM/S) measurements. STS allows to determine the size and uniformity of the superconducting gap across the sample surface. Although STS measurements cannot directly probe the substrate-film interface, it can detect changes in the film properties at the sample surface, which arise from differences at the substrate–film interface, similar to the observed changes in surface morphology (Fig.~\ref{fig1b} and ~\ref{stm_topo}).

Prior to STM/S measurements the samples were degassed \textit{in-situ} at 280\,\degree C for 20\,min to remove surface adsorbates. Cleaning via ion bombardment was avoided to minimize the risk of surface defect formation. Subsequently, the samples were transferred into a Createc scanning tunneling microscope and cooled down to 1.2\,K. All STM/S experiments were performed with an Ag-coated PtIr tip.

STM topographs are shown in Appendix \ref{App_STM_Topography} and are in broad agreement with the AFM data displayed in Fig.~\ref{fig1b}c and d.
Figure~\ref{spectra}a shows differential conductance spectra measured across representative areas of the B-Al$_2$O$_3$ and R31-Al$_2$O$_3$ sample respectively. A linear background was subtracted from each spectrum. From the STS data, three main conclusions can be drawn. Firstly, both the gap width and coherence peak height were not uniform along the measured lines. Clear variations in both parameters were observed on a length scale of 20 to 50\,nm. This observation is consistent with previous STS measurements on homogeneously disordered superconducting thin films \cite{Kamlapure2013, Dynes1986}. Secondly,in-gap states were observed, particularly on the R31-Al$_2$O$_3$ sample. Such sub-gap spectral weight has also been observed in STM measurements of TiN and may arise from local variations in tunnelling conditions, surface disorder, or Andreev contributions at highly transparent tip–sample contacts \cite{Liao2019}. Thirdly, some spectra showed unstable tunnelling, which results in a noisy signal and little to no superconducting gap. We attribute this to possible surface contamination caused by exposure of the films to air.

In Fig.~\ref{spectra}b, the average differential conductance spectra of the data shown in Fig.~\ref{spectra}a are plotted together with the corresponding fits of the Bardeen-Cooper-Schrieffer (BCS) density of states (DOS) \cite{tinkham}:

\begin{equation} \label{bcs_dos}
n_\mathrm{BCS}(E) =
n_0 \, \mathrm{Re} \left[ \frac{\lvert E \rvert}{\sqrt{E^2 - \Delta_{\rm BCS}^2}} \right],
\end{equation}

\noindent where $n_0$ is the normal state DOS, $E$ is the energy and $\Delta_{\rm BCS}$ is the superconducting pairing gap. In our experiment, we assumed a flat tip DOS and a constant tunneling matrix element, which allows us to use a simplified relation between the measured $dI/dV$ and $n_\mathrm{BCS}$:

\begin{equation} \label{didv_fit}
dI/dV
\propto \int_{-\infty}^{\infty}
\, dE \,
\left[ \frac{d}{dV} \, f_\mathrm{tip}(E)\right]
n_\mathrm{BCS}(E+eV),
\end{equation}

\noindent where $e$ is the elementary charge and $f_\mathrm{tip}(E)=1/[1 + \exp(E/k_\mathrm{B} T_\mathrm{tip})$, the Fermi-Dirac distribution function of the tunneling tip. From the fits in Fig.~\ref{spectra}b, pairing gaps of $\Delta_{\rm BCS}=(0.63\pm0.01)$ and $(0.67\pm0.01)\rm\,meV$ were obtained for the  B-Al$_2$O$_3$ and R31-Al$_2$O$_3$ samples respectively. Note that, while the majority of the spectra from Fig.~\ref{spectra}a are well described by the BCS scenario, the BCS gaps resulting from Fig.~\ref{spectra}b are averaged values that also include spectra with in-gap states. Such states broaden the averaged $dI/dV$ spectra, which can explain the increased effective temperatures of $T_{\rm tip}\approx1.90$\,K resulting from the fits. To clarify this issue, we plot the energy distributions of the coherence peak maxima in Fig.~\ref{spectra}c. The histograms emphasise that there are a few outliers corresponding to spectra with in-gap states, while the majority of spectra has coherence peaks centered around the mean values of $\tilde{\Delta}_{\rm coh}= 0.91$\,meV and 0.94\,meV for the B-Al$_2$O$_3$ and R31-Al$_2$O$_3$ sample respectively. By fitting these histograms to a Gaussian function:
\begin{equation} \label{gaussian_fit}
g(\Delta) \propto \exp\left( -\frac{1}{2} \frac{(\Delta_{\rm coh} - \tilde\Delta_{\rm coh}) ^2} {\sigma_{\rm coh} ^2}  \right),
\end{equation}
we further extract the width of the coherence peak distributions variations as $\sigma_{\rm coh}=0.02\rm\,meV$ for both samples. The small differences in the coherence peak positions are within observed variations measured across different locations on the samples. 

From the fitted order parameters, we calculate the BCS gap ratio $2\Delta_{\rm BCS}(0)/k_{\rm B}T_{\rm c}$. For sample B-Al$_2$O$_3$, we obtain 2.88, and for sample R31-Al$_2$O$_3$, 3.04. Both values are below the weak-coupling BCS value of 3.53. Nevertheless, we believe that our samples are well described by the BCS scenario, as this discrepancy can be explained by spatial inhomogeneities in our films: Fits of the averaged tunneling spectra in Fig.~\ref{spectra}b reflect the average order parameter across the surface. In contrast, $T_c$ determined from transport measurements corresponds more closely to the temperature below which a supercurrent can flow through interconnected regions with the largest order parameter. Based on the variations of the coherence peak positions shown in Fig.~\ref{spectra}c, we expect the variations in the order parameter throughout the film to be on the order of $\sim$10\,\%. Within this variation, the gap ratio approximately reaches the expected BCS value of 3.53. Remaining discrepancies may stem from STS being more sensitive to defects located at the film surface, especially in the film's native oxide layer. Such defects can locally weaken superconductivity, and therefore reduce the order parameter of the tunneling spectra (see Fig.~\ref{spectra}a) and thus the BCS gap ratio.

From these results, we conclude from the STS measurements that the pairing gap of the TiN films grown on both B-Al$_2$O$_3$ and R31-Al$_2$O$_3$ are comparable.
No significant influence of the thermal reconstruction of R31-Al$_2$O$_3$ on the pairing gap of TiN was observed, resulting in TiN films grown on both B-Al$_2$O$_3$ and R31-Al$_2$O$_3$ possessing similar superconducting properties.

\subsection{TiN CPW Resonators}
Finally, superconducting CPW resonators were fabricated from both TiN films in order to assess if the differences observed in crystalline quality with similar superconducting properties can effect the properties of said resonators. 

\begin{figure}[!h]
    \includegraphics[width =\linewidth]{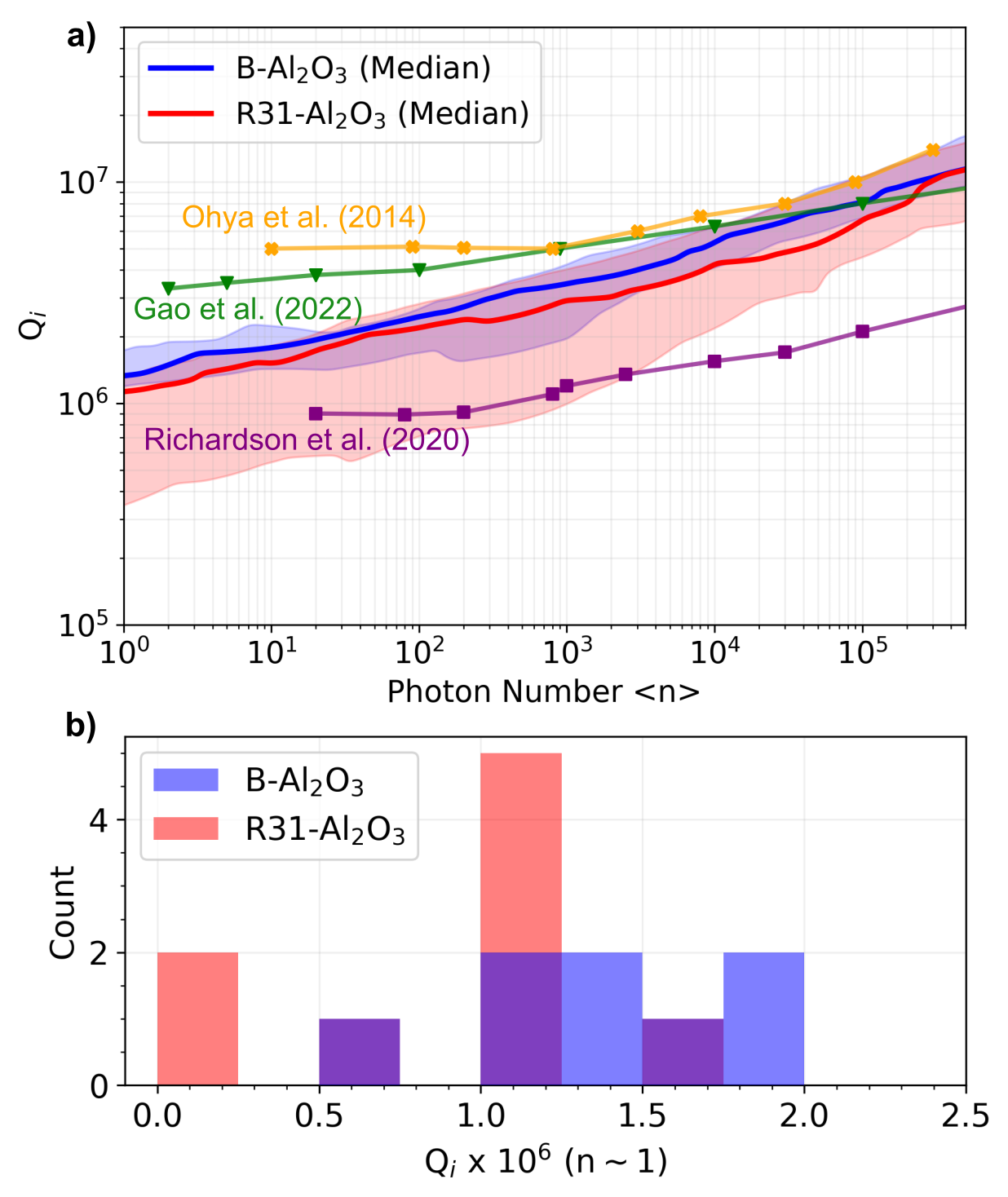}
    \caption{a) Median internal quality factor ($Q_i$) of TiN CPW resonators fabricated from $d$ = 106\,nm TiN films grown on R31-Al$_2$O$_3$ and B-Al$_2$O$_3$ respectively. The median value was only calculated from resonators where single-photon data was available. A confidence interval of one standard deviation for $Q_i$ values across all resonators on each chip is also indicated by the corresponding colored envelope. Selected $Q_i$ values from literature are also shown, with the values obtained here being similar to those achieved previously \cite{Ohya2014,Richardson2020,Gao2022}. b) Histogram of single photon $Q_i$ values for all resonators on each chip.}
    \label{resonators} 
\end{figure}

A full list of the parameters of all resonators on each chip are given in Appendix  \ref{App_ResonatorParameters}. From the raw $Q_i$ data for all resonators on each chip, a median $Q_i$ as a function of photon number, $\braket{n}$ was calculated and is indicated in Fig.~\ref{resonators}a, with a confidence interval equal to one standard deviation as a measure of device-to-device performance for resonators on a given chip. The raw $Q_i$ data for all resonators on each chip are shown in Appendix \ref{App_ResonatorParameters}, with the confidence interval overlaid.  Figure~\ref{resonators}b shows the distribution of $Q_i$ values at $\braket{n} \sim 1$ across all resonators, for which single-photon data was available, on each chip. For both chips, a majority of resonators had $Q_i$ values exceeding 10$^6$, similar to what has been observed in prior literature on TiN resonators \cite{Richardson2020,Ohya2014, Gao2022}. Median $Q_i$ values of the TiN resonators at single photon values were calculated at 1.33 $\times$ 10$^6$ and 1.13 $\times$ 10$^6$ for B-Al$_2$O$_3$ and R31-Al$_2$O$_3$ respectively. For each chip, the confidence interval at single photon levels spanned from 1.19 $\times$ 10$^6$ - 1.73 $\times$ 10$^6$ and 0.35 $\times$ 10$^6$ - 1.31 $\times$ 10$^6$ for B-Al$_2$O$_3$ and R31-Al$_2$O$_3$ respectively. All measured resonators exhibited a characteristic dip in its corresponding, frequency-dependent transmission function $S_{21}$, and the expected circular profile in complex space, examples of which on each chip are shown in Appendix \ref{App_ResonatorParameters} at single photon levels with corresponding fits \cite{Probst2015,Kalil2012}. The corresponding parameters for each resonator on both B-Al$_2$O$_3$ and R31-Al$_2$O$_3$ are also given in Appendix \ref{App_ResonatorParameters}. For both chips, several resonators possessed $Q_i$ values exceeding 10$^7$ at high photon numbers, matching values measured in similar TiN resonators in prior literature \cite{Ohya2014,Gao2022, Leduc2010}.

\section{Discussion}
Both TiN films grown on R31-Al$_2$O$_3$ and B-Al$_2$O$_3$ exhibit epitaxial growth of the TiN (111) phase, as evidenced by the XRD spectra shown in Fig.~\ref{fig2a}. Compared to the TiN film on B-Al$_2$O$_3$, the film grown on R31-Al$_2$O$_3$ demonstrates superior crystalline order and a reduced mosaic spread of TiN grains. This suggests that the atomically ordered surface of R31-Al$_2$O$_3$ facilitates coherent rotational alignment of the TiN film, thereby minimizing grain misorientation. Furthermore, the atomically-smooth terrace surface of R31-Al$_2$O$_3$ promotes single-domain epitaxy, enabling the formation of large grains and step-flow growth. This is supported by the replication of the underlying terrace structure of the R31-Al$_2$O$_3$ substrate in the deposited TiN film \cite{Smink2024,KimNitridesTLE2025,Fu2023}. The $T_c$ values of these TiN films were measured at 5.11\,K and 5.08\,K on R31-Al$_2$O$_3$ and B-Al$_2$O$_3$, respectively. These values fall below the state-of-the-art value of $T_c \sim 5.7$\,K recorded for thin films grown via magnetron sputtering or thermal laser epitaxy \cite{KimNitridesTLE2025,Zhang2025}. Firstly, the TiN films presented in this work were deposited at 1150\degree C, a temperature at which the rate of nitrogen diffusion out of the deposited film becomes significant \cite{Schulberg1996}. This stands in contrast to the growth of TiN via magnetron sputtering which is typically performed at significantly lower substrate temperatures or from stoichiometric targets \cite{Zhang2025,Ohya2014,Gao2022}. High temperature deposition would then result in small changes in the stoichiometry of the TiN, potentially via the formation of N vacancies, which would then reduce the $T_c$ of the film compared to bulk values. In order to minimise the potential presence of N vacancies and potentially improve the stoichiometry of TiN films grown at high substrate temperatures, the use of an Ammonia (NH$_3$) atmosphere during growth may be desirable, as demonstrated in prior literature demonstrating the growth of highly crystalline superconducting TMNs with state-of-the-art-properties \cite{KimNitridesTLE2025}.

Despite the clear advantages in crystal quality when growing TiN on R31-Al$_2$O$_3$, TiN resonators on R31-Al$_2$O$_3$ and B-Al$_2$O$_3$ exhibit similarly high $Q_i$ values, with low-power $Q_i$ values greater than 10$^6$ for many resonators. As shown in Fig. \ref{resonators}, this performance is consistent with other TiN resonators from literature, including TiN films grown via MBE, \cite{Richardson2020} or those grown via conventional sputtering methods \cite{Ohya2014,Gao2022}. However, we caution that directly comparing our work with prior literature may be misleading due to differing resonator geometries, growth mechanisms and choices in substrate material, all of which impact the extracted values of $Q_i$. In addition, we note that some prior literature may only report the measured $Q_i$ values from a small number of resonators, whereas the median $Q_i$ values presented in this work are calculated across all resonators on each chip.

A greater variation in resonator performance is observed for TiN resonators grown on R31-Al$_2$O$_3$ compared to B-Al$_2$O$_3$. 
However, the larger spread of low-power $Q_i$ values observed for resonators on R31-Al2O3 does not appear to correlate directly with the improved crystalline quality of the TiN film. In particular, the lowest-$Q_i$ devices on the R31 chip occur among resonators with the narrowest CPW geometry ($w$ = 3\,$\mu$m, $s$ = 2\,$\mu$m), for which the electric-field participation of the metal–vacuum and substrate–vacuum interfaces is enhanced \cite{Melville2020}. At single-photon powers, losses in TiN CPW resonators are known to be strongly influenced by TLS associated with processed surfaces and interfaces, and recent studies indicate that fabrication conditions and interfacial oxide chemistry can affect microwave loss more strongly than the crystalline orientation of the TiN film itself \cite{Melville2020, Calusine2018, Lang2026}. Consequently, local variations introduced during lithography, dry etching, resist removal or subsequent surface oxidation could produce appreciable device-to-device variation without being reflected in the XRD, transport or STS properties of the film. With the present number of devices and one chip for each substrate preparation method, we therefore cannot assign the broader $Q_i$ distribution uniquely to the R31 reconstruction. The larger laterally correlated structural domains observed on R31-Al$_2$O$_3$ may additionally reduce spatial averaging of local loss centres, but testing this interpretation would require correlating individual resonator locations with local morphology \cite{Chen2026}.

Regardless of this variation, both TiN films are capable of supporting resonators with similarly high $Q_i$ values. When taken together, we can conclude that thermal reconstruction of the Al$_2$O$_3$ substrate is a viable method of substrate preparation for superconducting devices with comparable quality to those grown on substrates which were prepared chemically. Indeed, this resilience of TiN resonator performance despite the differences in substrate preparation is broadly consistent with prior literature suggesting that dielectric losses within a superconducting quantum device are mostly dominated by metal-vacuum interfaces, which changes in the interface between the substrate and the film having minimal effect on device performance \cite{Woods2019,Altoe2022,Murthy2025}.

\section{Conclusion}
In summary, we have demonstrated that thermally reconstructed sapphire is capable of producing highly crystalline, epitaxial films grown by MBE, that have greater crystallinity than those grown on sapphire substrates prepared by aggressive chemical cleaning. We attribute this to the formation of an atomically-smooth terrace structure inherent to the R31 reconstruction that has been extensively studied and found to aid in the growth of epitaxial thin films with enhanced microstructure \cite{Smink2024,SmartSiTLE2024,MajerTa2024}.
While TiN films grown on thermally reconstructed sapphire exhibited lower mosaicity and greater long-range order, these films possessed similar superconducting properties to those grown on chemically prepared substrates. It may be possible that any deviations observed in the superconducting properties of our films compared to literature, may be due to the high temperature epitaxy of these TiN thin films on sapphire substrates, potentially leading to the formation of N vacancies. The quality of this high temperature epitaxial growth may be improved further via the use of Ammonia (NH$_3$) as a reactive atmosphere, potentially enhancing crystallinity further while minimising the formation of vacancies \cite{KimNitridesTLE2025}.
When fabricated into superconducting CPW resonators, both substrate preparation methods produce resonators with quality factors over 10$^6$ at single photon values, with some exceeding 10$^7$ at high photon numbers. No significant or systematic differences were observed in the superconducting properties or behavior of the resonators on either chip. This implies that thermal reconstruction of sapphire substrates via direct laser heating is a viable method of $in-situ$ substrate preparation that by-passes aggressive chemical cleaning. Thermal reconstruction of the substrate promotes greater crystalline coherence of the epitaxially grown film while preserving the electrical properties of the deposited material. This may also present exciting opportunities for the $in-situ$ growth of quantum devices and high-quality 2D materials on oxide substrates \cite{Hanna2025}. When taken together, these results establish thermally reconstructed sapphire as a robust \textit{in-situ} platform for the epitaxial growth of superconducting nitrides, opening new avenues for the fabrication of quantum devices and other heterostructures.

\begin{acknowledgments}
All chemical preparation and fabrication steps detailed within this work were performed within the Helmholtz Nano Facility, to which we are grateful to all the technical staff \cite{albrecht2017helmholtz}. We wish to specifically highlight the technical support of  J\'{e}ferson R. Guimar\~{a}es, Christoph Krause and Anja Zaß.
The authors thank Marc Scheffler, Yayi Lin, Christine Falter and Nils von den Driesch for many insightful scientific discussions. The authors thank Lucas Bothe for assistance with RHEED operation and Amin Karimi for establishing the STM experimental setup and for his assistance with the STM measurements. The authors gratefully acknowledge support by the German Federal Ministry of Education and Research (BMBF), funding program "Quantum technologies - from basic research to market", project QSolid (Grant No. 13N16149) and from the European Union Flagship on Quantum Technology (OpenSuperQplus100, No. 101113946).
\end{acknowledgments}

\section*{Conflict of Interest}
The authors declare no competing interests.

\section*{Data Availability}
All of the data from this work is available upon reasonable request from the authors.

\section*{Author Contributions}
\textbf{Thomas J. Smart}: Data curation (equal); Conceptualization (lead); Formal analysis (equal); Investigation (lead); Methodology (equal); Software (supporting); Validation (equal); Visualisation (lead); Writing – original draft (lead); Writing – review \& editing (equal).
\textbf{Marc Neis}: Data curation (equal); Conceptualization (equal); Formal analysis (equal); Investigation (equal); Methodology (equal); Software (lead); Validation (equal); Visualisation (equal); Writing – original draft (supporting); Writing – review \& editing (equal).
\textbf{Janine Lorenz}: Data curation (equal); Formal analysis (equal); Investigation (supporting); Methodology (equal); Validation (equal); Visualisation (equal); Writing – original draft (supporting); Writing – review \& editing (supporting).
\textbf{Marcello P. Guardascione}: Data curation (supporting); Investigation (equal); Methodology (equal); Validation (supporting).
\textbf{Roudy Hanna}: Data curation (supporting); Investigation (supporting); Methodology (equal); Validation (supporting).
\textbf{Michael Schleenvoigt}: Investigation (equal); Methodology (equal); Writing – review \& editing (supporting).
\textbf{Yuan Gao}: Methodology (supporting); Validation (supporting); Software (supporting).
\textbf{Joscha Domnick}: Investigation (equal); Methodology (equal). 
\textbf{Benjamin Bennemann}: Methodology (equal).
\textbf{Abdur Rehman Jalil}: Formal Analysis (equal).
\textbf{Jin Hee Bae}: Data curation (equal).
\textbf{Harsh Bhardwaj}: Methodology (supporting); Software (supporting).
\textbf{F. Stefan Tautz}: Funding acquisition (equal); Supervision (equal).
\textbf{Felix Lüpke}: Methodology (supporting), Investigation (supporting), Supervision (equal).
\textbf{Detlev Grützmacher}: Conceptualization (equal); Funding acquisition (equal); Supervision (equal).
\textbf{Rami Barends}: Funding acquisition (equal); Project administration (equal); Supervision (equal); Writing – original draft (equal); Writing – review \& editing (equal).
\textbf{Pavel A. Bushev}: Conceptualization (equal); Funding acquisition (equal); Project administration (equal); Supervision (equal); Writing – original draft (equal); Writing – review \& editing (equal).
\textbf{Peter Schüffelgen}: Conceptualization (equal); Funding acquisition (equal); Supervision (equal); Writing – review \& editing (equal).

\appendix

\setcounter{figure}{0}  
\renewcommand{\thefigure}{\Alph{section}\arabic{figure}}  

\onecolumngrid
\section{\label{RHEED_Appendix} RHEED images of R31-Al$_2$O$_3$}

As described in the main text, $10 \times 10$\,mm $c$-plane $\alpha$-Al$_2$O$_3$ substrates were heated at a rate of 3\,°C/s to 1700\,°C, where they were held at this temperature for 200\,s before being cooled back to room temperature at a rate of 3\,°C/s. During heating, the TLE chamber pressure remained in the order of 10$^{-9}$\,mbar. These substrates were stored in air for six months after reconstruction.
In order to verify the presence of a specific double-step ($\sqrt{31}$$\times$$\sqrt{31}$)\,R$\pm$9\degree (R31) surface reconstruction, Reflection High-Energy Electron Diffraction (RHEED) was performed on the samples at 15\,keV. Before imaging, the samples were baked in a UHV atmosphere at 400\,°C for 30\,min to remove any volatile absorbates (eg. water vapor) resulting from extended storage in air. This recovers the original surface reconstruction, which in the case of R31-Al$_2$O$_3$ is stable in air \cite{Smink2024}.

\begin{figure}[!h]
    \includegraphics[width =\linewidth]{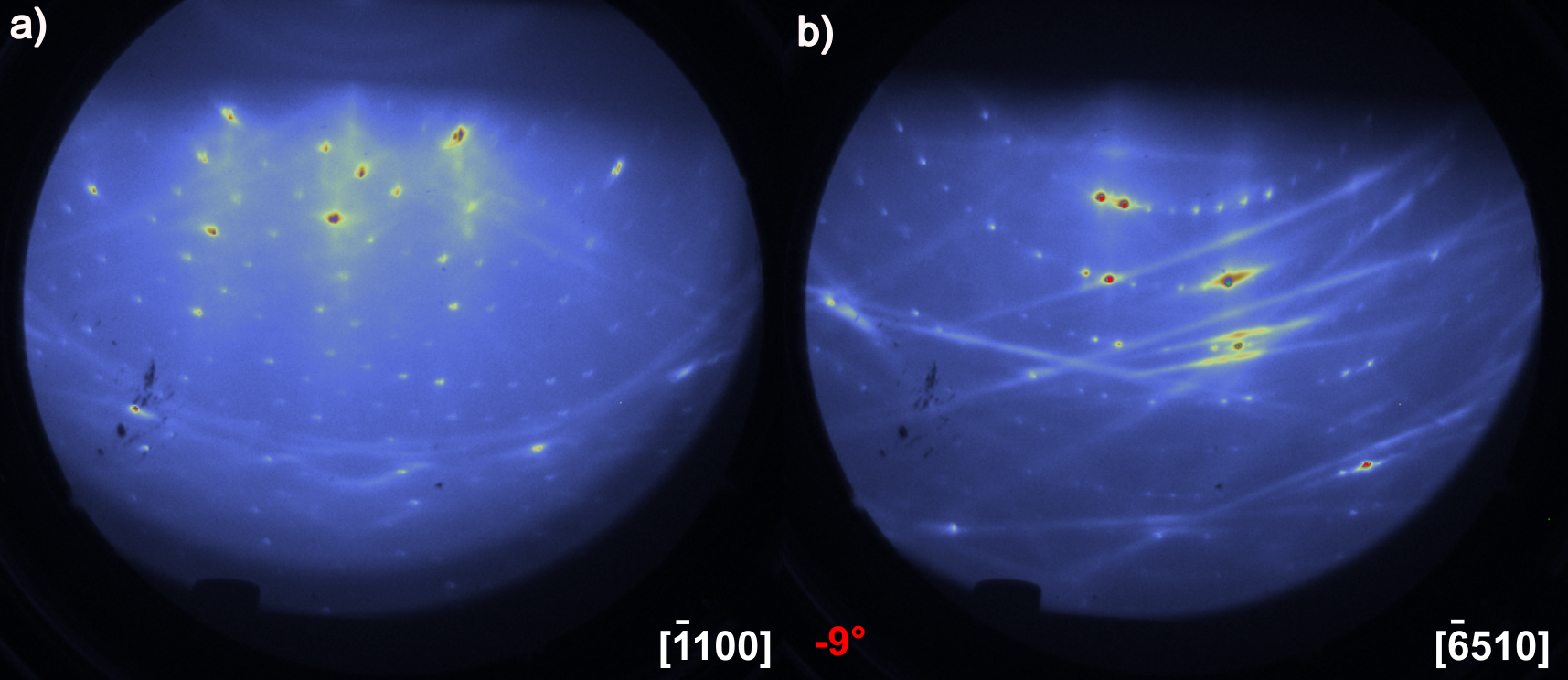}
    \caption{RHEED images of the reconstructed sapphire surface produced after being heated to 1700\,°C for 200\,s in UHV conditions. Prior to imaging, these substrates were stored in air for six months. The sample was baked in a UHV atmosphere at 400\,°C for 30\,min prior to imaging in order to remove any volatile absorbates resulting from extended storage in air. Images were taken along the [$\bar{1}$100] direction, shown in a) and the [$\bar{6}$510] direction, taken by rotating the sample by 9° anti-clockwise (while viewing the substrate from the top). This image is shown in b). Both of these images are consistent with prior literature of the double-step ($\sqrt{31}$$\times$$\sqrt{31}$)\,R$\pm$9\degree surface reconstruction \cite{Smink2024}.}
    \label{Fig-RHEED} 
\end{figure}

As shown in Fig. \ref{Fig-RHEED}, both images present sharp and well-resolved spots, implying a highly ordered 2D surface, consistent with prior literature of the reconstructed R31 surface \cite{Tosaka2014, Smink2024}. Within Fig. \ref{Fig-RHEED}, all spots appear to lie upon observable Laue circles centred on the upper-right corner of the image. By rotating the pattern by -9°, corresponding to an anti-clockwise rotation, as viewed from above the substrate, the concentric circular pattern in Fig. \ref{Fig-RHEED}b is observed. Crucially, this pattern in Fig. \ref{Fig-RHEED}b is not present when the sample is rotated by +9°, indicating that the degeneracy of the ($\sqrt{31}$$\times$$\sqrt{31}$)\,R$\pm$9\degree reconstruction has been lifted, supporting a double-step R31 surface reconstruction with a singular orientation of +9\degree. 
The observation that the RHEED pattern observed in Fig. \ref{Fig-RHEED} was recovered, despite storage in air for six months supports prior literature stating that the R31 reconstruction remains stable in air and does not degrade \cite{Smink2024}. When taken as a whole, these images confirm the presence of a stable R31 reconstruction after direct laser heating of the Al$_2$O$_3$ substrates at 1700\degree C for 200\,s. The lack of degeneracy in the RHEED pattern when rotating the RHEED pattern to $\pm$9\degree from the [$\bar{1}$100] azimuth supports the formation of a double-step surface with a singular surface rotation of +9\degree, consistent with the AFM image of the same R31-Al$_2$O$_3$ surface shown in Fig.\ref{fig1b}.

\section{\label{App_PoleFigures} TiN/Al$_2$O$_3$ XRD Pole Measurements}

To confirm the presence of a fixed epitaxial relation between the c-plane $\alpha$-Al$_2$O$_3$ and the TiN (111) films grown on top of them, XRD pole measurements centred on TiN(111) and off-axis Al$_2$O$_3$(10$\bar{1}$4) reflection for both the B-Al$_2$O$_3$ and R31-Al$_2$O$_3$ substrates.

\begin{figure}[!h]
    \includegraphics[width =\linewidth]{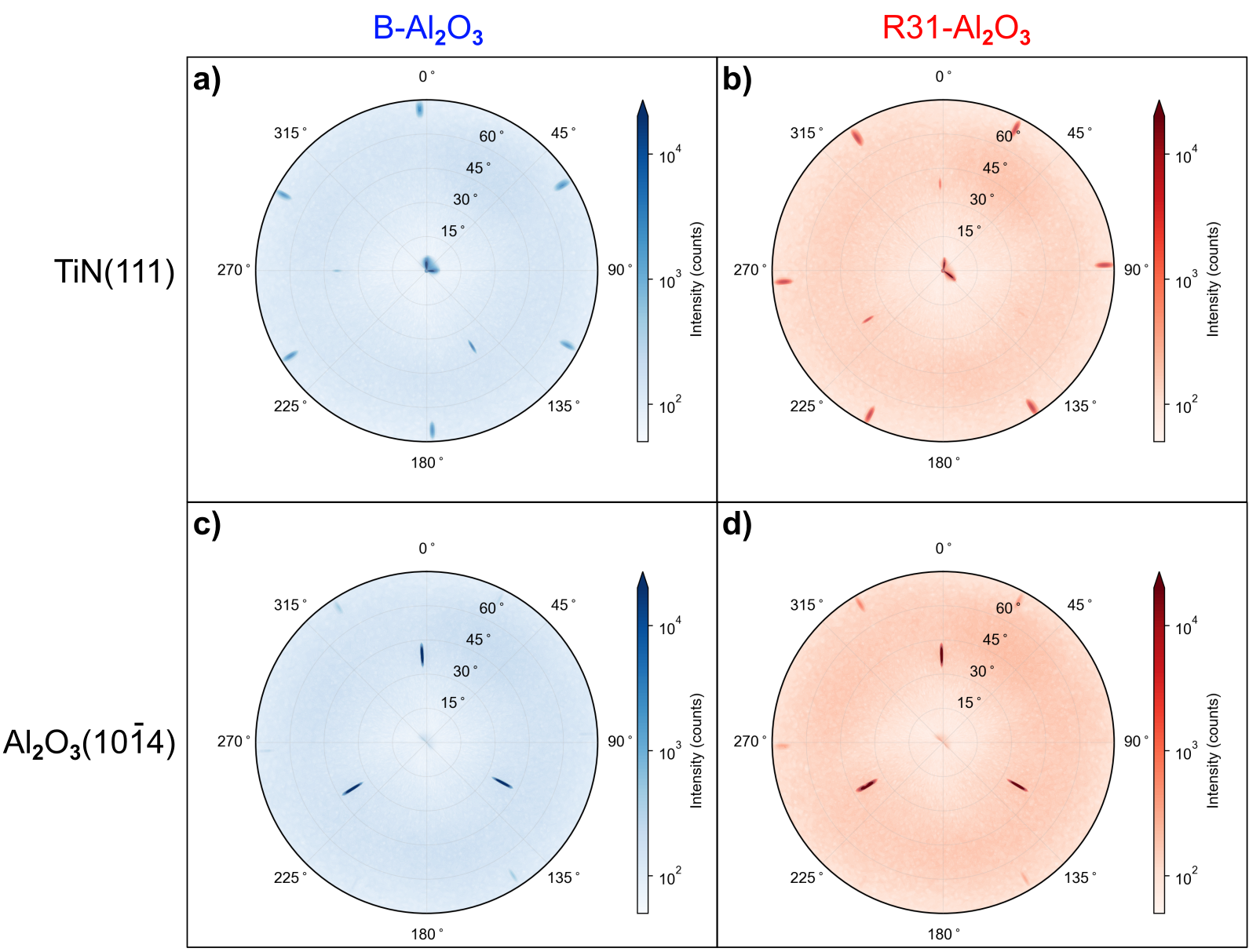}
    \caption{XRD pole figures centred around the TiN(111) and off-axis Al$_2$O$_3$ (10-14) peaks, for TiN films on R31-Al$_2$O$_3$ and B-Al$_2$O$_3$.}
    \label{Fig-PoleFigure} 
\end{figure}

Pole-figure measurements shown in Fig. \ref{Fig-PoleFigure} reveal TiN grows epitaxially with
TiN (111) $\parallel$ Al$_2$O$_3$(0001). The TiN {111} pole figures indicate six discrete off-axis poles at $\chi$ $\sim$ 70.5\degree, consistent with two 60°-related in-plane rotational variants. Both the reconstructed and unreconstructed sapphire substrates yield TiN (111) films with the same epitaxial orientation relationship relative to the underlying Al$_2$O$_3$ lattice. Although the absolute azimuthal positions of the pole sets differ by approximately 30\degree between samples, the TiN and Al$_2$O$_3$ poles rotate together, leaving their relative azimuthal relationship unchanged.

\section{\label{App_SC-Properties} Superconducting Properties of TiN films}

Before the TiN films on B-Al$_2$O$_3$ and R31-Al$_2$O$_3$, room temperature four-point probe measurements were performed using a needle probe station to determine the resistivity, $\rho$ of the samples. These values were determined to be 21.1\,$\mu\Omega\cdot cm$ and  19.8\,$\mu\Omega\cdot cm$ for TiN on B-Al$_2$O$_3$ and R31-Al$_2$O$_3$ respectively. 
\begin{figure}[!h]
    \includegraphics[width =\linewidth]{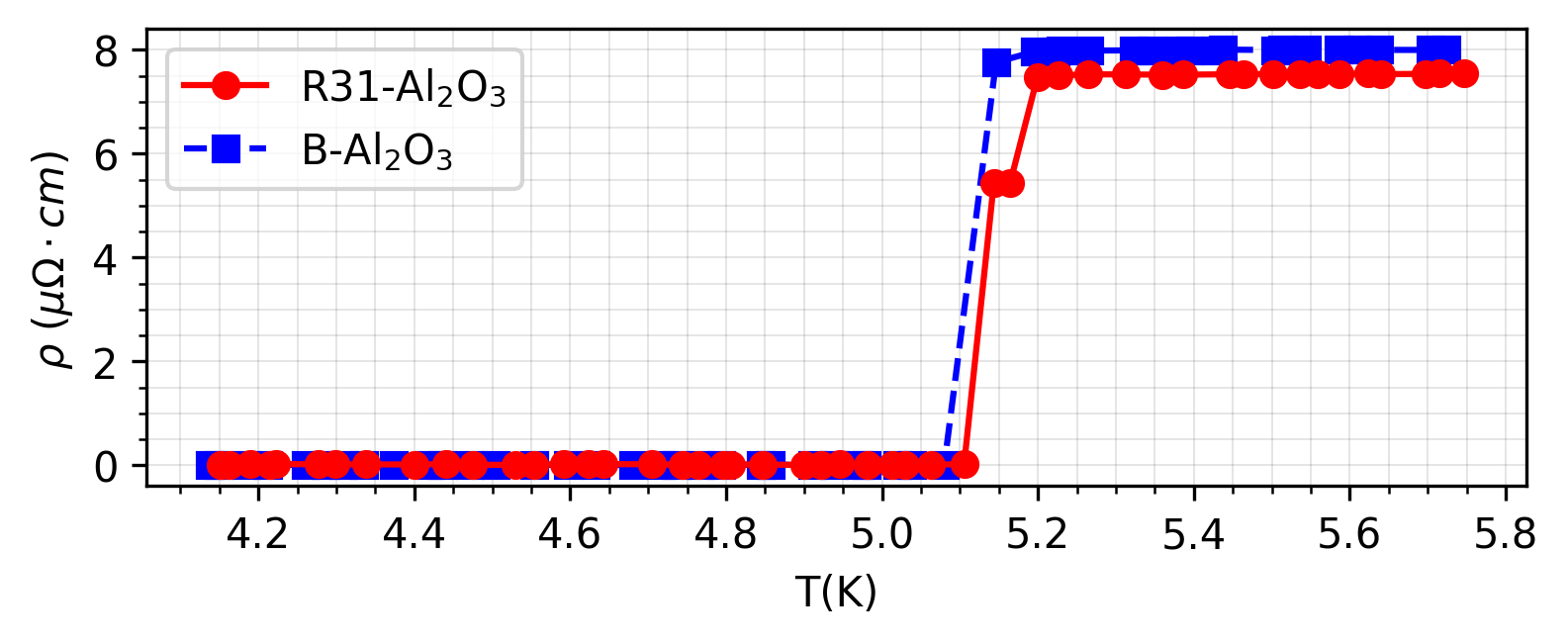}
    \caption{$\rho$(T) around the region of $T_c$ for $d$= 106\,nm thick TiN films grown on B-Al$_2$O$_3$ and R31-Al$_2$O$_3$ substrates.}
    \label{Tc-comp} 
\end{figure}
These samples were then placed within a dilution refrigerator with $\rho$ being measured as a function of temperature. From the results in Fig. \ref{Tc-comp}, the $T_c$ of each TiN film was determined to be 5.08\,K and 5.11\,K on B-Al$_2$O$_3$ and R31-Al$_2$O$_3$ respectively. The value of $T_c$ was determined as the temperature at which the superconducting transition was complete when the samples were cooled from their normal state to the superconducting state. Due to the limited resolution of the $T_c$ measurement, it is challenging to assign a width, $\Delta T_c$ to the superconducting transition of both films. An extreme, approximate value of $\Delta T_c$ can be assigned to both films accounting for the region over which the entire superconducting transition occurs. This results in $\Delta T_c$ = 0.07\,K and $\Delta T_c$ = 0.09\,K for TiN on B-Al$_2$O$_3$ and R31-Al$_2$O$_3$ respectively.

\section{\label{App_STM_Topography}STM Topography}

\begin{figure}[!h]
    \includegraphics[width =\linewidth]{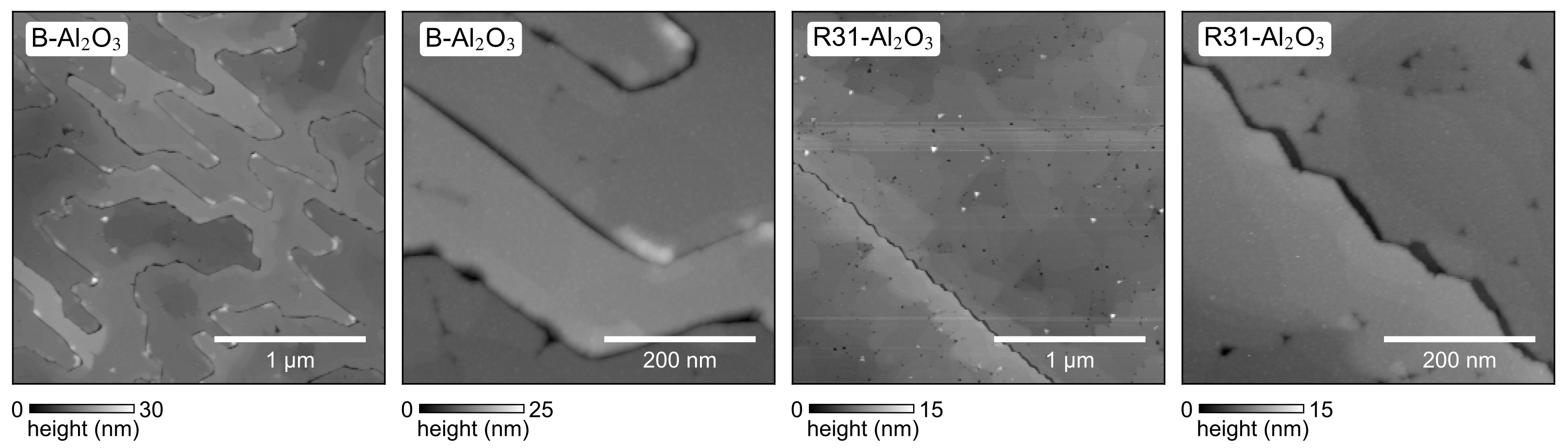}
    \caption{STM topographs of $d$= 106\,nm thick TiN films grown on chemically cleaned (B-Al$_2$O$_3$) and thermally reconstructed (R31-Al$_2$O$_3$) sapphire substrates (current setpoint: 30\,pA, bias voltage: 800\,mV).}
    \label{stm_topo} 
\end{figure}

Fig. \ref{stm_topo} depicts representative STM topographs of both samples. 
We attribute differences in the height variations between the STM and AFM data to the AFM having a blunter tip and thus lower lateral resolution, which leads to an underestimation of the surface roughness.

\section{\label{App_ResonatorParameters}Resonator Parameters}

\begin{figure}[!h]
    \includegraphics[width =\linewidth]{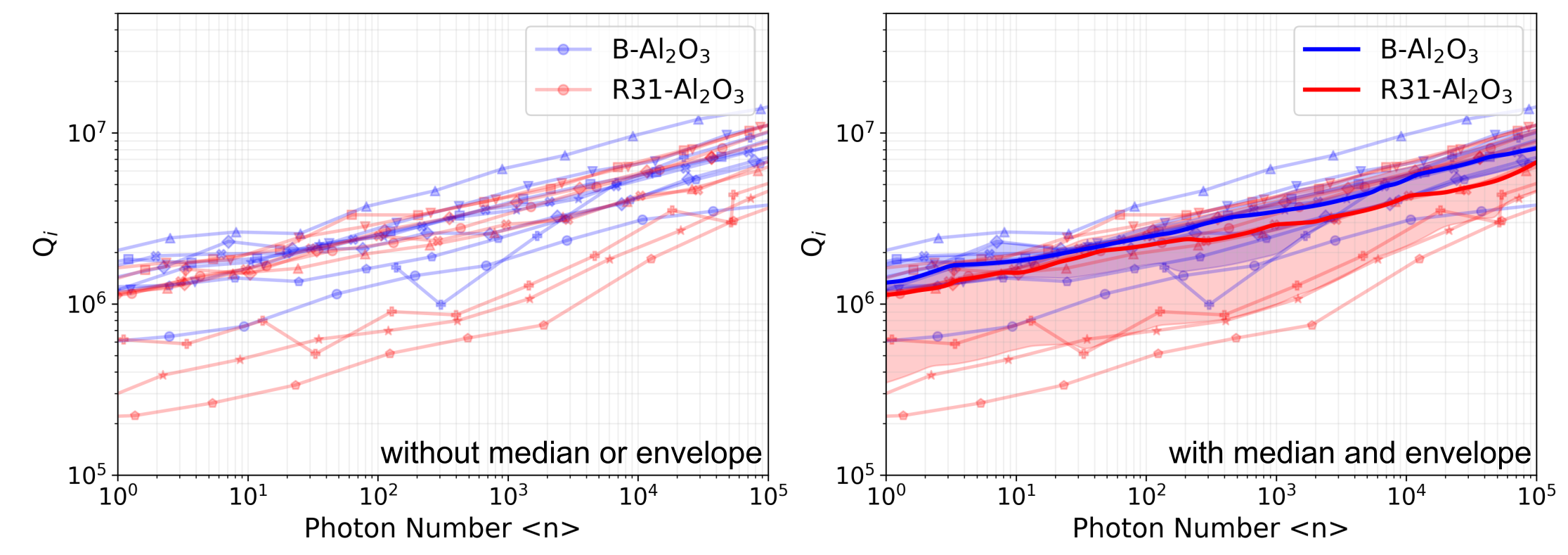}
    \caption{The first panel shows the $Q_i$ data for all CPW resonators fabricated from $d$ = 106\,nm TiN films on R31-Al$_2$O$_3$ and B-Al$_2$O$_3$ respectively. Each resonator is shown with a different marker and is colored according to which chip they belong to. The second panel includes the raw data overlaid with a median Q$_i$ value for each chip. A confidence interval is also shown for each chip. This confidence interval corresponds to one standard deviation from the mean Q$_i$ value.}
    \label{resonators-raw} 
\end{figure}

Table \ref{table:1} and Table \ref{table:2} show the resonator parameters for all TiN resonators prepared on both R31-Al$_2$O$_3$ and B-Al$_2$O$_3$ respectively. For each resonator, both the internal quality factor, $Q_i$ and coupling quality factor, $Q_c$ were measured at low microwave powers, with the corresponding value of $\braket{n}$ indicated. All resonators possessed a coupler length of $l$ = 50\,\textmu m. The raw $Q_i$ data as a function of $\braket{n}$ of all measured resonators on each chip, along with the extracted median $Q_i$ and confidence interval of each chip are shown in Fig. \ref{resonators-raw}. 

\begin{table}[h]
\caption{Resonator parameters for $d$ = 106\,nm TiN film on R31-Al$_2$O$_3$ at given photon number $\braket{n}$}
\label{table:1}
\begin{tabular}{|r|r|r|r|r|r|r|} 
 \hline
 Frequency (GHz) & Microwave Power (dBm) & $\braket{n}$ & $Q_i$ ($\times$ 10$^6$) &  $Q_c$ ($\times$ 10$^6$) & s (\textmu m)  & w (\textmu m) \\  
 \hline
 4.818 & -155  & 1.63  & 1.581 & 0.367 & 2  & 3\\
 \hline
 5.777 & -150 & 1.28 & 1.149 & 0.106 & 6 & 10\\
\hline
 5.872 & -155 & 0.71 & 1.582 & 0.207 & 2  & 3\\
\hline
 6.797 & -150 & 0.76 & 1.156 & 0.085 & 6 & 10\\
\hline
 6.890 & -150 & 1.11 & 0.618 & 0.185 & 8 & 12\\
\hline
 6.941 & -145 & 1.35 & 0.218 & 0.626 & 2 & 3\\
\hline
7.846 & -150 & 0.89 & 1.123 & 0.146 & 6 & 10\\
\hline
7.946 & -150 & 0.96 & 1.125 & 0.167 & 8 & 12\\
\hline
8.083 & -150 & 0.38 & 0.383 & 0.222 & 2 & 3\\
\hline
\end{tabular}
\end{table}

\begin{table}[h]
\caption{Resonator parameters for $d$ = 106\,nm TiN film on B-Al$_2$O$_3$ at given photon number $\braket{n}$}
\label{table:2}
\begin{tabular}{|r|r|r|r|r|r|r|} 
 \hline
 Frequency (GHz) & Microwave Power (dBm) & $\braket{n}$ & $Q_i$ ($\times$ 10$^6$) &  $Q_c$ ($\times$ 10$^6$) & s (\textmu m) & w (\textmu m)  \\
 \hline
 4.960 & -155  & 1.21  & 1.826 & 0.254 & 8 & 12\\
 \hline
 5.340 & -155 & 0.71 & 0.593 & 1.143 & 2 & 3\\
\hline
 5.762 & -150 & 1.24 & 1.216 & 0.101 & 6 & 10\\
\hline
5.863 & -150 & 2.52 & 1.953 & 0.207 & 8 & 12\\
\hline
5.971 & -135 & 138.41 & 1.630 & 0.617 & 2 & 3\\
\hline
6.827 & -150 & 0.76 & 1.254 & 0.083 & 6 & 10\\
\hline
6.908 & -155 & 0.57 & 1.512 & 0.243 & 2 & 3\\
\hline
7.845 & -150 & 0.65 & 1.344 & 0.096 & 6 & 10\\
\hline
7.938 & -150 & 0.96 & 1.198 & 0.164 & 8 & 12\\
\hline
\end{tabular}
\end{table}

\begin{figure}[h]
    \includegraphics[width =0.8\linewidth]{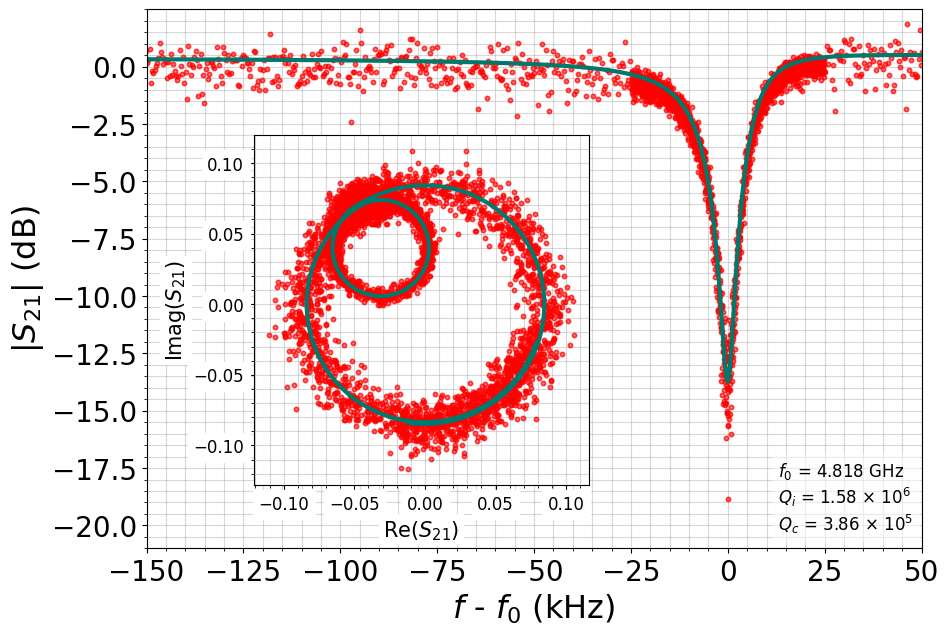}
    \caption{$S_{21}$ as a function of frequency, $f$ for a TiN resonator on R31-Al$_2$O$_3$ with a resonant frequency, $f_0$ = 4.818\,GHz. All data was taken at $T$ $\sim$ 20\,mK and measured at a microwave power of -155\,dBm corresponding to $\braket{n}$ $\sim$ 1.68.}
    \label{Resonator-R31} 
\end{figure}

\begin{figure}[h]
    \includegraphics[width =0.8\linewidth]{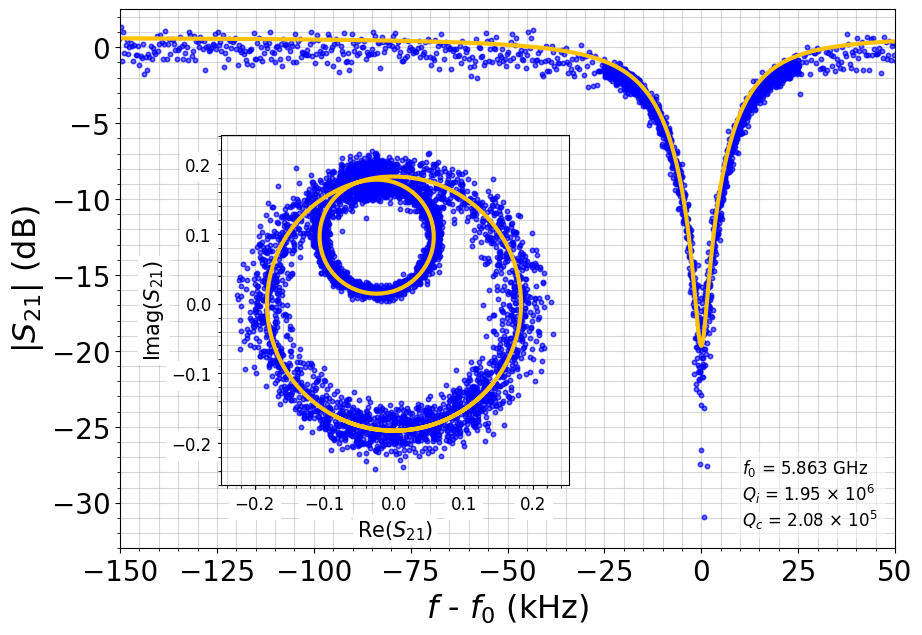}
    \caption{$S_{21}$ as a function of frequency, $f$ for a TiN resonator on B-Al$_2$O$_3$ with a resonant frequency, $f_0$ = 5.863\,GHz. All data was taken at $T$ $\sim$ 25\,mK and measured at a microwave power of -150\,dBm corresponding to $\braket{n}$ $\sim$ 2.52.}
    \label{Resonator-B} 
\end{figure}

\newpage
\paragraph*{Example S$_{21}$ fits} In Fig. \ref{Resonator-R31} and Fig. \ref{Resonator-B}, raw $S_{21}$ data measured at single photon levels from a given TiN CPW resonator are shown. Data from resonators on R31-Al$_2$O$_3$ and B-Al$_2$O$_3$ are included. To extract the properties in question a Laurentzian fit for the normalized inverse transmission $\tilde{S}_{21}^{-1}$ was used \cite{Megrant2012}:

\begin{equation}
\tilde{S}_{21}^{-1} = 1 + \frac{Q_{i}}{Q_{c}^{*}} e^{i\phi} \frac{1}{1 + 2iQ_{i}\delta x}
\end{equation}

With the internal quality factor, $Q_i$, the rescaled coupling quality factor $Q_{c}^{*} = (Z_0 / |Z|)$, impedance mismatch $\phi$, and the linear frequency detuning from the resonance frequency $\delta x = (f - f_0)/f_0$. Additionally the resonator is corrected for various external electrical factors. The average photon number of the resonators $\braket{n}$ was calculated via \cite{BarendsThesis}:
\begin{equation}
    \braket{n} = \frac{4}{\pi hf_{r}} \cdot CP_{in}Z_{in} \frac{Q_l^2}{Q_c}
\end{equation}

using the total resonator capacitance, $C$, feedline power, $P_{in}$, feedline impedance $Z_{in}$ and the loaded quality factor $Q_l = \frac{Q_iQ_c}{Q_i + Q_c}$.

\clearpage

\twocolumngrid
\bibliography{references}

\end{document}